\documentclass[onecolumn,floatfix,aps,nofootinbib]{revtex4}
\usepackage[dvips]{epsfig}
\usepackage[english]{babel}
\usepackage{bbm}
\usepackage{verbatim}
\usepackage{array}
\usepackage{bm} 
\usepackage{amsmath}
\usepackage{amsfonts}
\usepackage{amssymb}
\usepackage{hyperref}
\usepackage[dvipsnames]{xcolor}
\hypersetup{colorlinks=true,linkbordercolor=Blue,linkcolor=Blue, citecolor=Blue}
\usepackage{indentfirst} 
\usepackage[titletoc]{appendix}

\newcommand{\e}{\!\!\;}

\usepackage{bbold}

\begin{document}

\title{Strongly continuous representations in the Hilbert space: a far-reaching concept}

\author{J. M. Hoff da Silva} 
\email{julio.hoff@unesp.br}
\affiliation{Departamento de F\'isica, Universidade
Estadual Paulista, UNESP, Av. Dr. Ariberto Pereira da Cunha, 333, Guaratinguet\'a, SP,
Brazil.}
\author{G. M. Caires da Rocha} 
\email{gabriel.marcondes@unesp.br}
\affiliation{Departamento de F\'isica, Universidade
	Estadual Paulista, UNESP, Av. Dr. Ariberto Pereira da Cunha, 333, Guaratinguet\'a, SP,
	Brazil.}



\begin{abstract}
	
	We revisit the fundamental notion of continuity in representation theory, with special attention to the study of quantum physics. After studying the main theorem in the context of representation theory, we draw attention to the significant aspect of continuity in the analytic foundations of Wigner's work. We conclude the paper by reviewing the connection between continuity, the possibility of defining certain local groups, and their relation to projective representations.   
	
\end{abstract}		

\maketitle

\section{Introduction}

Since the bygone days of the 1930s, Wigner has collected strong mathematical results on mappings of continuous groups on a given Hilbert space \cite{wig1}.  From this time comes the famous theorem that ensures that any symmetry operation (any operation preserving probabilities) can be represented by either a linear and unitary or antilinear and antiunitary operator (see Appendix). The pursuit of Poincar\`e group representations on Hilbert spaces, a program supported by Weyl \cite{weyl}, culminates in the profound and robust work celebrated here \cite{wig2}.   

It is a complicated, if not impossible, task to pick out the most effective results from the wealth of excellent ones shown in \cite{wig2}. Wigner introduced in this work the very precise notion of particle, to mention only one of the main achievements. However, the mathematical partial results collected on the way to such relevant insights are also important. We would like to revisit one of these cornerstone results in the existence of so-called admissible representatives, presenting its meaning based on continuous projective representations. Let us introduce some standard notions.

The first idea to remember is the quantum mechanical ray representation of states, which arises from the freedom that a complex field gives to states in the Hilbert space inner product. The transition probability for a quantum state $\psi$ to turn into $\phi$ is expressed by $P = |(\phi,\psi)|^{2}$. This probability remains the same if we replace the states by rays, $\bm{\Psi}$. Each ray comprises an equivalence class of states containing all elements $\{\alpha\psi\}$ (where $\alpha$ is a unimodular complex phase), of which a given element $\alpha\psi\in\bm{\Psi}$, where $\psi$ is fixed, is a representative. The inner product between two rays is defined as $\bm{\Phi}\cdot \bm{\Psi}=|(\phi,\psi)|$, where $\phi\in\bm{\Phi}$ and $\psi\in\bm{\Psi}$. In a manner akin to vector rays, Bargmann introduced the concept of operator rays \cite{bar} $\textbf{U}$ as the set comprising all elements $\alpha U$ with fixed $U$ and a unimodular complex $\alpha$. 

In the case where the symmetries of a physical system are described by a given continuous group $H$, there exists an isomorphism between each element $h\in H$ and an operator ray $\textbf{U}_h$ such that for $h_1,h_2\in H$, the usual representation relation (although related to operator rays) holds, i. e. $\textbf{U}_{h_1}\textbf{U}_{h_2}=\textbf{U}_{h_{12}}$. These are the projective representations whose continuity will be studied here. In physics, of course, continuous representations are of great importance. In a prosaic context, for example, the regular textbooks of Quantum Mechanics and Quantum Field Theory introduce the possibility of acting on quantum states (e.g., with differential operators) without regard to possible mathematical subtleties. This and many others aspects are automatically taken as well-posed mathematically in a standard exposition of quantum physics. Thanks to the work of Wigner, in particular the aforementioned theorem, the approach presented in these textbooks is not wrong. 

To justify our choice from a different framework, as as Wigner himself recognized \cite{wig2}, part of the generality and novelty of his approach came from mathematical rigor, in particular concerning continuity. Majorana \cite{maj} and Dirac \cite{dir} dopted continuity in several aspects of their analysis (an approach which is certainly justifiable and entirely correct, after all) but whose dubiety was completely removed by Wigner.  

 Now back to the (projective) representations frame: from a mathematical point of view, phases in representations of continuous groups have become quite a sophisticated tool with the aid of algebraic topology \cite{neeb} since the early presentation due to Bargmann. More or less recently, the original line of research gained additional interest by generalizations of the phase exponents (which enter the projective representations) involving the time (or the spacetime) parameter(s) \cite{rec,consec}. In this generalization of Bargmann's theory, the selection of admissible representatives is also an important issue.  

This review is devoted to an appreciation of conditions related to the existing argument of regular representatives, whose steps are also revisited. We begin in Section II with a recapitulation of the theorem that ensures the selection of a strongly continuous set of ray operators representatives in the Hilbert space, and discuss its standard proof (Wigner-Bargmann) step by step, discussing various aspects of it. Important enough, the existence of such representatives is at the very heart of our understanding of the quantum process, from spacetime evolution to probability transitions. In Section III we call special attention to some relevant consequences of continuity. Of course, we would not be so far saying, somewhat frivolously, that everything comes from continuity. There are, nevertheless, profound concepts linked to continuity that can be brought up in the analysis and whose full appreciation is usually bypassed. In Section III.A we define and investigate the continuity of local factors and study its implications to representation theory. Besides, in this section, we also place a theorem related to one-parametric group representation quite relevant in the Wigner approach. In Section III.B we explore how continuity also helps in the understanding of whether a given representation is projective or genuine. In the final section, we conclude. To guarantee a sequential reading of the paper we leave for the Appendix a step by step proof of Wigner's famous theorem about symmetric representations in the Hilbert space, as well as another result, relevant for the action upon connected (sub)groups. 

This work is a modest contribution to the subject. We do not present new results, but give a technical appreciation to the foundations of representation theory, explicitly presenting all the proof steps and highlighting the physical interest of the results when the situation appears. Whenever possible we follow Bargmann's exposition \cite{bar} by its didactic and far-reaching results connecting continuity to physical aspects. Throughout the paper, operators will be taken as unitary (see Proposition 5 in the Appendix) and vector states are normalized to unity.     

\section{Selection of continuous representatives}

We begin by defining standard tools for representations of continuous groups. Let $\mathcal{H}$ be a complex Hilbert space. Let the distance, $d(\bm{\Psi},\bm{\Psi''})$, between two rays $\bm{\Psi}$ and $\bm{\Psi''}$ belonging to $\mathcal{H}$ be given by the minimum value of $||\psi-\psi''||$, where double bar stands for the usual vector norm $||\psi||=(\psi,\psi)^{1/2}$ and $\psi$ and $\psi''$ are representatives of $\bm{\Psi}$ and $\bm{\Psi''}$, respectively. This qualitative definition can be fully specified by noting that $||\psi-\psi'' ||^2=2(1-Re(\psi,\psi''))$, from which it may be proved that   
\begin{equation}\label{distancia}
d(\bm{\Psi},\bm{\Psi''})
=
[2(1-\bm{\Psi}\cdot\bm{\Psi}'')]^{\frac{1}{2}},
\end{equation}

\noindent where the definition of the inner product between two rays has already been defined in the introduction. To see that Eq. (\ref{distancia}) holds, take $\bm{\Psi}=\sigma\psi$ and $\bm{\Psi''}=\tau\psi''$, with $|\sigma|=1=|\tau|$. It is always possible to find a unimodular complex number, say $\lambda$, such that $(\psi,\psi'')=\lambda|(\psi,\psi'')|$. So it is clear that 
\begin{eqnarray}\label{nov}
||\psi -\psi''||^2=2(1-Re[\sigma^*\tau(\psi,\psi'')]),
\end{eqnarray} where $\sigma^*$ stands for the complex conjugation of $\sigma$. Note that 
\begin{eqnarray}
|(\psi,\psi'')|=|\sigma||\tau||(\psi,\psi'')|=|\sigma^*||\tau||(\psi,\psi'')|=|\sigma^*\tau(\psi,\psi'')|=|(\sigma\psi,\tau\psi'')|=\bm{\Psi}\cdot\bm{\Psi''}
\end{eqnarray} and therefore we can set $(\psi,\psi'')=\lambda\bm{\Psi}\cdot\bm{\Psi''}$. Returning to (\ref{nov}), we are left with ($\bm{\Psi}\cdot\bm{\Psi''}\in \mathbb{R}$)
\begin{equation}
||\psi -\psi''||^2=2(1-Re[\sigma^*\tau\lambda]\bm{\Psi}\cdot\bm{\Psi''}).\label{novnov}
\end{equation} It is clear that $||\psi -\psi''||^2$ reaches its minimum for $Re[\sigma^*\tau\lambda]_{\max}=1$, from which Eq. (\ref{distancia}) follows. As we will see in a moment, the definition of distance is crucial for the very conception of (strong) continuity for ray representations. 
\medbreak
{\bf Definition 1:} {\it A given ray representation of a (Lie) group $H$ is said to be continuous if for any element $h\in H$, any $\bm{\Psi}\in \mathcal{H}$, and any $\epsilon>0$, there exists a neighborhood $\mathfrak{N}\subset H$ of $h$ such that $d(\textbf{U}_s\bm{\Psi},\textbf{U}_h\bm{\Psi})<\epsilon$, if $s\in \mathfrak{N}$.}      
\medbreak
As an aside remark, we note that it suffices to consider the above definition for the identity element. Moreover, in a complete (metric) space, $A$ is said to be continuous with respect to $B$ if the set of nonzero values of $A$ is bounded by $B$. In this sense, we can say that the inner product in both terms is continuous with respect to the distance $d(\bm{\Psi},\bm{\Psi''})$. Here is the proof: suppose four rays $\bm{\Psi}_{1}$, $\bm{\Psi}_{2}$, $\bm{\Phi}_{1}$ and $\bm{\Phi}_{2}$ belong to $\mathcal{H}$. Then 
\begin{eqnarray}\label{}
|\bm{\Psi}_{1}\cdot\bm{\Phi}_{1}-\bm{\Psi}_{2}\cdot\bm{\Phi}_{2}|
&=&\left.
|
\bm{\Psi}_{1}\cdot\bm{\Phi}_{1}
-
\bm{\Psi}_{1}\cdot\bm{\Phi}_{2}
+
\bm{\Psi}_{1}\cdot\bm{\Phi}_{2}
-
\bm{\Psi}_{2}\cdot\bm{\Phi}_{2}
| \right.\nonumber
\\
&\leq&\left.
|
(\psi_{1},\phi_{1})
-
(\psi_{1},\phi_{2})
|
+
|
(\psi_{1},\phi_{2})
-
(\psi_{2},\phi_{2})
|=
||\phi_{1}-\phi_{2}|| + ||\psi_{1}-\psi_{2}||,\right.
\qquad\qquad
\end{eqnarray} and therefore (since the inequality always holds) we have

\begin{equation}
|\bm{\Psi}_{1}\cdot\bm{\Phi}_{1}-\bm{\Psi}_{2}\cdot\bm{\Phi}_{2}|\leq d(\bm{\Phi}_{1},\bm{\Phi}_{2}) + d(\bm{\Psi}_{1},\bm{\Psi}_{2}).
\end{equation}  

Now we will state and discuss in detail the theorem asserting the selection possibility of admissible representatives. 
\medbreak

{\bf Theorem 1:} {\it Let $\textbf{U}_{r}$ be a continuous ray representation of a group $G$. For all $r$ in a suitably chosen neighborhood $\mathfrak{N}_{0}$ of the unit element $e$ of $G$,  one can select a strongly continuous set of representatives $U_{r}\in \textbf{U}_{r}$ (i.e., for any vector $\psi$, any $r\in \mathfrak{N}_{0}$ and any positive $\epsilon$, there exists a neighborhood $\mathfrak{N}$ of $r$ such that $||U_{s}\psi - U_{r}\psi||<\epsilon$ if $s\in\mathfrak{N}$).} 
	
\medbreak
The set of representatives $\{U_{r}\}$ satisfying these conditions is called an admissible set of representatives. From the concept of an admissible set of representatives derives a wealth of important results in representation theory in physics (ultimately including the particle concept itself). This set is indeed strongly continuous and probability transitions taken from its representatives vary continuously with the group element $s$, in complete agreement with the previous definition. We will discuss in detail the proof of Bargmann, who in turn followed the Wigner steps.  
\medbreak
{\bf Proof of Theorem 1:} Let $\bm{\Psi}$ be a fixed ray in $\mathcal{H}$ and $\psi$ a given representative. Define $g_r=\bm{\Psi}\cdot\textbf{U}_r\bm{\Psi}$ for $r\in H$. since the inner product is continuous, $g_r$ is a continuous function of $r$. Therefore, it is possible to choose a suitable neighborhood $\mathfrak{N}\supset r$ such that $\alpha<g_r\leq 1$ with $\alpha\in (0,1)$. Moreover, as a strategy for the proof, a certain representative $U_r \in \textbf{U}_r$ is chosen such that 
\begin{equation}
g_r=\bm{\Psi}\cdot\textbf{U}_r\bm{\Psi}=(\psi,U_r\psi).\label{j1}
\end{equation} Note the absence of modulus\footnote{This choice was made by Wigner \cite{wig2}, p. 169.} in (\ref{j1}), which is contrary to the definition of the inner product of vector rays. We will address this point after completing the standard proof. Until then, we'll only emphasize that $e\in\mathfrak{N}$, just as it is contained in the statement.

Let $\psi\in\bm{\Psi}$, $r,s\in\mathfrak{N}$ and define the quantities (partially preserving the Bargmann notation) 
\begin{equation}\label{j2}
d_{r,s}(\psi)=d(\textbf{U}_{r}\bm{\Psi},\textbf{U}_{s}\bm{\Psi}),
\end{equation}
\begin{equation}\label{j3}
\sigma_{r,s}(\psi)=(U_{r}\psi,U_{s}\psi),
\end{equation}
\begin{equation}\label{j4}
Z_{r,s}(\psi)=U_{s}\psi-\sigma_{r,s}(\psi)U_{r}\psi.
\end{equation} These quantities will help the proof process. Note that $Z_{r,s}(\psi)$ is orthogonal to $U_r\psi$, as can be easily seen from 
\begin{eqnarray}
(U_{r}\psi,Z_{r,s}(\psi))=(U_{r}\psi,U_{s}\psi)-\sigma_{r,s}(\psi),
\end{eqnarray} which vanishes by means of (\ref{j3}). From this, one can see that 
\begin{eqnarray}
||Z_{r,s}(\psi)||^2=(U_{s}\psi-\sigma_{r,s}(\psi)U_{r}\psi,U_{s}\psi-\sigma_{r,s}(\psi)U_{r}\psi)=1-\sigma_{r,s}(\psi)(U_s\psi,U_r\psi)
\end{eqnarray} and therefore (again using (\ref{j3}))
\begin{eqnarray}
||Z_{r,s}(\psi)||^2=1-|\sigma_{r,s}(\psi)|^2.\label{jj}
\end{eqnarray} It follows straightforwardly from Eq. (\ref{distancia}) that $1-|(\psi,\phi)|^2\leq d^2$, leading to
\begin{equation}
||Z_{r,s}(\psi)||^2\leq d^2_{r,s}(\psi).\label{j5}
\end{equation} Now, taking $\psi=\phi$ and calculating $(\phi,Z_{r,s}(\phi))$, we have 
\begin{equation}
\sigma_{r,s}(\phi)=\frac{1}{g_r}[g_s-(\phi,Z_{r,s}(\phi))]. \label{j6}
\end{equation} Recalling that $||\psi-\psi'' ||^2=2(1-Re(\psi,\psi''))$ we have\footnote{In fact, calling $(\psi,\phi)=x+iy$ with $x,y \in\mathbb{R}$, we see that $[1-Re(\psi,\phi)]_{\max}=(1-x)_{\max}\leq [(1-x)_{\max}^2+y^2]^{1/2}$ from which the inequality follows.} $||\psi-\psi'' ||^2\leq 2|1-(\psi,\psi'')|$. Therefore 
\begin{eqnarray}
||U_s\phi-U_r\phi||^2\leq 2|1-(U_r\phi,U_s\phi)|=2|1-\sigma_{r,s}(\phi)| \label{nec}
\end{eqnarray} and by (\ref{j6}) we find
\begin{equation}
||U_s\phi-U_r\phi||^2\leq 2 \Big|\frac{1}{g_r}[g_r-g_s+(\phi,Z_{r,s}(\phi))]\Big|.\label{j7}
\end{equation} Since $g_r>\alpha$, it is possible to rewrite the above equation as 
\begin{eqnarray}
||U_s\phi-U_r\phi||^2\leq \frac{2}{\alpha}|g_r-g_s+(\phi,Z_{r,s}(\phi))|\leq \frac{2}{\alpha} [|g_r-g_s|+|(\phi,Z_{r,s}(\phi))|].\label{j8}
\end{eqnarray}
 
As defined before, the functions $g_r$ give $|g_r-g_s|=|(\phi,U_r\phi)-(\phi,U_s\phi)|=|(\phi,U_r\phi-U_s\phi)|\leq ||U_r\phi-U_s\phi||$. Therefore, $|g_r-g_s|$ is less than (or equal to) any value of $||U_r\phi-U_s\phi||$; in particular the inequality holds for the minimum value of $||U_r\phi-U_s\phi||_{\min}=d_{r,s}(\phi)$. Therefore $|g_r-g_s|\leq d_{r,s}(\phi)$. Moreover, $|(\phi,Z_{r,s}(\phi))|\leq ||Z_{r,s}(\phi)||$ and using (\ref{j5}) we have $|(\phi,Z_{r,s}(\phi))|\leq d_{r,s}(\phi)$. Collecting all these results, we finally get 
\begin{equation}
||U_s\phi-U_r\phi||^2\leq \frac{4}{\alpha}d_{r,s}(\phi),\label{jjj}  
\end{equation} which ensures continuity for $U_r\phi$ in the sense of the highlighted definition before the theorem. The next step is to ensure continuity for a vector $\chi$ given by $\chi=(\phi+\varphi)/\sqrt{2}$, where the normalized vector $\varphi$ is assumed to be orthogonal\footnote{The reader may here appreciate the inventiveness of Wigner's approach: in general considerations about Quantum Mechanics the dimension of the underling Hilbert space is not {\it a priori} specified. It is indeed so, since it is the quantum mechanical problem that dictates the dimension. This simple observation reveals the finesse of Wigner's procedure in proving the continuity for $\phi$ and $\chi$.} to $\phi$. Note that 
\begin{equation}
(U_r\phi,Z_{r,s}(\chi))=(U_{r}\phi,U_{s}\chi)-\sigma_{r,s}(\chi)(U_{r}\phi,U_{r}\chi),\label{jjj4}
\end{equation} now adding and subtracting $(U_{s}h,U_{s}k)$ to (\ref{jjj4}), the result may be recast as 
\begin{equation}
(U_r\phi,Z_{r,s}(\chi))=(U_{r}\phi-U_{s}\phi, U_s\chi)+(U_s\phi,U_s\chi)-\sigma_{r,s}(\chi)(U_{r}\phi,U_{r}\chi). \label{jjj1}
\end{equation} In turn, $(U_m\phi,U_m\chi)=(\phi,U_m^\dagger U_m\chi)=(\phi, \frac{1}{\sqrt{2}}[\phi+\varphi])=1/\sqrt{2}$ for every $m\in H$, in particular for $m\in\mathfrak{N}\subset H$. Returning to (\ref{jjj4}) then we have
\begin{equation}
	(U_r\phi,Z_{r,s}(\chi))+(U_s\phi-U_r\phi,U_s\chi)=\frac{1}{\sqrt{2}}(1-\sigma_{r,s}(\chi)).\label{jjj5}
\end{equation} 	
  
Now we can apply Eq. (\ref{jjj5}) to (\ref{nec}) (suitable adequate to $\chi$) and arrive at 
\begin{equation}
||U_s\chi-U_r\chi||^2\leq 2^{2/3} \Big\{ |(U_r\phi,Z_{r,s}(\chi))|+|(U_s\phi-U_r\phi,U_s\chi)|\Big\}.\label{jjj6}
\end{equation} Both terms of the right-hand side of Eq. (\ref{jjj6}) are bounded from above by Schwarz inequality. Then, using Eq. (\ref{j5}) in the first term, we have  
\begin{equation}
||U_s\chi-U_r\chi||^2\leq 2^{2/3} \Big\{d_{r,s}(\chi)+||U_s\phi-U_r\phi||\Big\} \label{pre}
\end{equation} and the continuity of $U_r\phi$ implies the continuity of $U_r\chi$ (and of course of $U_r\varphi$). Finally, if $\psi$ is a linear combination written in terms of $\phi$ and $\varphi$, as defined earlier, then $U_r\psi$ is clearly continuous. \hspace{.3cm} $\Box$
\medbreak
As a final remark before going further, note that the first steps of the proof could be repeated around any group element $k\in H$ by simply adapting the neighborhood to include $k$ and starting with the definition of $g_r$ functions as $g_r=\textbf{U}_k\bm{\Psi}\cdot \textbf{U}_r(\textbf{U}_k\bm{\Psi})$.    

\subsection{Additional Discussion}

Apart from the comments on the proof inserted here and there, let us concentrate on the determination of the $g_r$ functions. First, a fact: there is no loss of generality in choosing a neighborhood of $e\in H$ such that $g_r>\alpha$ for $\alpha\in (0,1)$ or even in setting the operator such that $g_r=(\psi,U_r\psi)$ instead of $|(\psi,U_r\psi)|$. The continuity of the inner product with respect to the distance ensures this last procedure. However, it is instructive, to see the effect of such a device within the proof scheme when the necessity emerges, as it were, rather than {\it a priori}.  

Let us begin with functions $\tilde{g}_r=\bm{\Psi}\cdot\textbf{U}_r\bm{\Psi}=|(\psi,U_r\psi)|$ for which the group separability condition $\tilde{g}_r>\alpha$ for $r\in\mathfrak{N}\subset H$ holds. Given this definition, part of the procedure used in the previous proof is unhelpful. We first note that with $Z_{r,s}(\phi)$, as defined in (\ref{j4}), 
\begin{eqnarray}
|(\phi,Z_{r,s}(\phi))|=|(\phi, U_s\phi)-\sigma_{r,s}(\phi)(\phi,U_r\phi)|\geq |(\phi,U_s\phi)|-|\sigma_{r,s}(\phi)||(\phi,U_r\phi)|,\label{ad1}  
\end{eqnarray} using the standard triangle inequality for complex numbers. In a more compact form
\begin{equation}
|(\phi,Z_{r,s}(\phi))|\geq \tilde{g}_s-|\sigma_{r,s}(\phi)|\tilde{g}_r.\label{ad2}
\end{equation} From (\ref{ad2}), we read 
\begin{equation}
|1-|\sigma_{r,s}(\phi)||\leq \frac{1}{\alpha} \Big\{|\tilde{g}_r-\tilde{g}_s|+|(\phi,Z_{r,s}(\phi)\e)|\Big\}.\label{ad3}
\end{equation} Again, both terms in the right-hand side of (\ref{ad3}) are bounded from above by the distance in the Hilbert space. In the second term, Eq. (\ref{j5}) will be used again, while for the first term we have
\begin{eqnarray}
|\tilde{g}_r-\tilde{g}_s|=||(\phi,U_r\phi)|-|(\phi,U_s\phi)||\leq |g_r-g_s|\label{ad4}
\end{eqnarray} and the discussion around (\ref{j8}) holds. Hence
\begin{eqnarray}
|1-|\sigma_{r,s}(\phi)||\leq \frac{2}{\alpha}d_{r,s}(\phi).\label{ad5}
\end{eqnarray} 

Now recall that a direct computation leads to Eq. (\ref{nec}), i.e.
\begin{equation}
|| U_s\phi-U_r\phi||^2\leq 2|1-\sigma_{r,s}(\phi)|
\end{equation} and hence the procedure to ensure continuity of $U_r\phi$ requires additional attention, since $|1-\sigma_{r,s}(\phi)|\geq |1-|\sigma_{r,s}(\phi)||$. Let us examine both cases. 

In the case that $|1-\sigma_{r,s}(\phi)|>|1-|\sigma_{r,s}(\phi)||$ one is not able to compare $|| U_s\phi-U_r\phi||^2$ and $d_{r,s}(\phi)$ accurately. This situation then leads to an empty tautology: the selection of continuous representatives is the one that choose $U_r\phi$ respecting $|| U_s\phi-U_r\phi||^2\leq (4/\alpha) d_{r,s}(\phi)$, i.e. the continuous one. 

However, when $|1-\sigma_{r,s}(\phi)|=|1-|\sigma_{r,s}(\phi)||$, one can indeed claim continuity for $U_r\phi$ as can be easily seen. This condition (leading to the proof) is mathematically satisfied whenever $Re(\sigma_{r,s}(\phi))\geq 0$ and $Im(\sigma_{r,s}(\phi))$ vanishes. But these conditions are precisely the conditions studied when setting $g_r$ functions (without tilde). Indeed
\begin{equation}
\sigma_{r,s}(\phi)=(U_r\phi,U_s\phi)=(\phi,U^\dagger_rU_s\phi)=(\phi,U_{r^{-1}s}\phi),\label{n1}
\end{equation} is nothing but a $g_m$ function for $m=r^{-1}s\in\mathfrak{N}$. Therefore, given the conditions $Re(g_m(\phi))\geq 0$ and $Im(g_m(\phi))=0$ we have $g_m=\tilde{g}_m$ leading to\footnote{The rest of the proof follows strictly the remaining steps of the proof performed in the preceding section.} $|| U_s\phi-U_r\phi||^2\leq (4/\alpha) d_{r,s}(\phi)$. Going further, by calling $U_m=\tau U^0_{m}$, $\phi=\delta\phi^0$ for fixed $U^0_m$ and $\phi^0$ and denoting $(\phi^0,U^0_m\phi^0)=X+iY$ and $\tau=\tau_1+i\tau_2$ ($X, Y, \tau_1, \tau_2 \in\mathbb{R}$), the conditions give     
\begin{eqnarray}
X\tau_1-Y\tau_2\geq 0,\nonumber\\
X\tau_2+Y\tau_1=0.\label{n2}
\end{eqnarray} It is not difficult to satisfy Eqs. (\ref{n2}) without any operator particularization. The last additional remark is that in the absence of a phase in the representation, i.e. for genuine (not projective) representations, continuity for the representatives is ultimately attainable by continuity of the inner product.

\section{Consequences of Continuity}

We will here point out some direct consequences of continuity, relevant to the mathematical structure underlining the understanding of quantum physics, firstly examining local factors and then investigating some consequences of the selection of continuous representatives together with freedom in the selection procedure. 

\subsection{Continuous local factors}

Let us assume implicitly in what follows that all group elements to be worked out in this section belong to the same neighborhood $\mathfrak{N}$ (or to suitable intersections of neighborhoods), so that the group operations are locally well-defined. This requires $e\in H$ to be an element of the neighborhood. An admissible set of representatives engenders a continuous ray representation in $\mathfrak{N}$ and since $U_rU_s$ and $U_{rs}$ belongs to the same ray, we have
\begin{equation}
U_rU_s=\omega(r,s)U_{rs},\label{n3}
\end{equation} where $|\omega(r,s)|=1$ and clearly $\omega(r,e)=\omega(e,s)=\omega(e,e)=1$, since $U_e=\mathbb{1}$. Note that the associative law of the group representation implies 
\begin{equation}
\omega(r,s)\omega(rs,m)=\omega(s,m)\omega(r,sm).\label{n4}
\end{equation} The functions $\omega(r,s)$ are the so-called local factors of a given ray representation and the continuity of admissible representatives leads to the continuity of them. Let us demonstrate this fact. 
\medbreak
{\bf Lemma 1:} {\it For an admissible set of representatives the local factors are continuous.}  
\medbreak

{\bf Proof of Lemma 1:} We start with a simple truism. Let $\psi\in\mathcal{H}$, then obviously  
\begin{eqnarray}
[\omega(r',s')-\omega(r,s)]U_{r's'}\psi=\omega(r',s')U_{r's'}\psi-\omega(r,s)U_{r's'}\psi.\label{n5}
\end{eqnarray} Now adding and subtracting the terms $U_{r'}U_s\psi$ and $U_rU_s\psi$ we obtain, using (\ref{n3}), 
\begin{eqnarray}
[\omega(r',s')-\omega(r,s)]U_{r's'}\psi=\omega(r,s)(U_{rs}-U_{r's'})\psi+U_{r'}(U_{s'}-U_s)\psi+(U_{r'}-U_r)U_s\psi. \label{n6}
\end{eqnarray} Thus we arrive at 
\begin{eqnarray}
||[\omega(r',s')-\omega(r,s)]U_{r's'}\psi||\leq ||\omega(r,s)(U_{rs}-U_{r's'})\psi||+||U_{r'}(U_{s'}-U_s)\psi||+||(U_{r'}-U_r)U_s\psi||. \label{n7}
\end{eqnarray} The left-hand side of (\ref{n7}) simplifies\footnote{In fact, for a complex $A$, $(AU_m\psi,AU_m\psi)^{1/2}=|A|(U_m\psi,U_m\psi)^{1/2}=|A|(\psi,U_{m^{-1}}U_m\psi)^{1/2}=|A|$ for normalized vectors.} to $|\omega(r',s')-\omega(r,s)|$. Moreover, since $|\omega(r,s)|=1$ and $U_r$ is unitary, the terms of the right-hand side are easy to handle. Thus, calling $U_{s'}\psi=\psi'$, we have 
\begin{equation}
|\omega(r',s')-\omega(r,s)|\leq ||(U_{rs}-U_{r's'})\psi||+||(U_{s}-U_{s'})\psi||+||(U_{r'}-U_{r})\psi'||,\label{n8}
\end{equation} from which the local factors for admissible representatives are indeed continuous. \hspace{.3cm} $\Box$
\medbreak
Continuity of phase factors can also be used to treat differentiability precisely \cite{bar}. We will just note here that a multidimensional Lie group $H$ has elements in bijective correspondence with open balls containing $\mathfrak{N}$ in an Euclidean space of same dimensionality \cite{naka}. To fix ideas, let us consider a fixed element $s$ and think of $r=r(r^1,r^2,\ldots,r^n)$, where $n=\dim(H)$ and $r^i$ $(i=1,\ldots,n)$, as the coordinates of $r$. The unity for the local factor may be written as $1=\omega(e,s)$ and the unity element $e\in H$ has coordinates given by $(0,0,\ldots,0)$. It follows that
\begin{eqnarray}
|\omega(r,s)-1|=|\omega(0+r^1, 0+r^2, \ldots, 0+r^n,s)-\omega(0,0,\ldots,0,s)|:\mathfrak{N}\subset \mathbb{R}^n\rightarrow\mathbb{R}^+.\label{n10}
\end{eqnarray} For the argument, if we take $s'=e=r'$ in (\ref{n8}) and denote $||(1-U_m)\psi||\leq\kappa d^{1/2}_{e,m}$ where $\kappa$ stands for a constant (a form certainly valid for admissible representatives), then (\ref{n8}) reads 
\begin{equation}
|\omega(0+r^1, 0+r^2, \ldots, 0+r^n,s)-\omega(0,0,\ldots,0,s)|\leq \kappa \sum_{i=r,s,rs} d_{e,i}^{1/2}\label{n9}
\end{equation} and the continuity of admissible representatives naturally bounds the local factor from above, in the sense just described.


So far we have said that the local factors are group elements dependent. It is easy to see that there is no dependence of local factors with respect to the state on which the group elements are represented \cite{wei}. Let $\psi_i$ $(i=1,2)$ be two linearly independent vectors in $\mathcal{H}$ and assume a dependence of the local phases on the states. Also assume $\psi_{3}=\psi_1+\psi_2$. It follows that $U_rU_s\psi_3=\omega_3(r,s)U_{rs}\psi_3$. However (recall that the operators acting in $\mathcal{H}$ are unitary and linear)
\begin{eqnarray}
\omega_3(r,s)U_{rs}(\psi_1+\psi_2)=\omega_1(r,s)U_{rs}\psi_1+\omega_2(r,s)U_{rs}\psi_2 \label{n11}
\end{eqnarray} and acting from the left in (\ref{n11}) with $U_{s^{-1}r^{-1}}$ we have $\omega_3(r,s)=\omega_1(r,s)=\omega_2(r,s)$, so no state dependence at all.   

Before further studying the local factors and their continuity, we make a parenthetical but important remark about continuity in irreductible representations and one-parametric Lie subgroups. A local one-parameter Lie subgroup is a continuous curve $g=g(\lambda)$ in $\mathfrak{N}$ ($\lambda$ takes values in an open real interval) with $g(\lambda_1)g(\lambda_2)=g(\lambda_1+\lambda_2)$ for $\lambda_1,\lambda_2,\lambda_1+\lambda_2$ defined in the same open interval in $\mathbb{R}$. In view of this remark we can state the following theorem \cite{wig2}.   
\medbreak
{\bf Theorem 2:} { \it Let $U_{g(\lambda)}$ be an infinitesimal operator of a continuous one-parametric subgroup. If there exists a physical state, say $\Psi$, on which the application of $U_{g(\lambda)}$ is well defined, then there exists an everywhere dense set of such states for irreducible representations.} 
\medbreak
{\bf Proof of Theorem 2:} If there exists such a $\Psi$ state, then the limit $\lim_{\lambda\rightarrow 0}\lambda^{-1}(U_{g(\lambda)}-1)\Psi$ is clearly well-defined. Now let $S$ be a given operator of the representation. The one-parametric property $U_{g(\lambda_1)}U_{g(\lambda_2)}=U_{g(\lambda_1+\lambda_2)}$ also holds for $S^{-1}U_{g(\lambda)}S$ and therefore the limit 
\begin{eqnarray}
\lim_{\lambda\rightarrow 0}\frac{1}{\lambda}(S^{-1}U_{g(\lambda)}S-1)\Psi \label{n12}
\end{eqnarray} is also well-defined. Taking $1=S^{-1}S$ in the above expression, we see that the following limit is also well-defined
\begin{equation}
\lim_{\lambda\rightarrow 0}\frac{1}{\lambda}(U_{g(\lambda)}-1)S\Psi. \label{n13}
\end{equation} This means that as far as the infinitesimal operator can be applied to $\Psi$, it can also be applied to $S\Psi$. Thus, for irreductible representations, the existence of $\Psi\in\mathcal{H}$, which implies (\ref{n13}), means that for every state on which infinitesimal $U_{g(\lambda)}$ can act, every $\mathfrak{V}\subset\mathcal{H}$ contains at least one $S\Psi$ that has exactly this property, giving rise to an everywhere dense set of such states in $\mathcal{H}$.  \hspace{.3cm} $\Box$  
\medbreak
Just as an additional remark about the theorem just proved, if $\psi$ does not belong to an irreducible representation then one cannot claim the existence of such an everywhere dense set (see \cite{wig2} for further discussion). Finally, this theorem states that infinitesimal operators in the Hilbert space can be treated in a somewhat ordinary way.    

\subsection{Exploring the continuity of representatives}

The continuity of representatives and local factors proved so far can be summarized in the expressions (\ref{pre}) and (\ref{n8}). It is now important to consider this strong continuity property together with the dependence of local factors on the choice of representatives. To begin with, consider $|\phi(r)|=1$ and select representatives in the same ray such that $U'(r)=\phi(r)U_r$, $\forall r\in \mathfrak{N} \subset H$. Of course, $U'(r)U'(s)=\omega'(r,s)U'(rs)$, which shows that the $\omega(r,s)$ function depends on the selection of representatives, given that
\begin{equation}
\omega'(r,s)=\omega(r,s)\frac{\phi(r)\phi(s)}{\phi(rs)}.\label{b1}
\end{equation} This freedom in the representative selection is obviously inherited from the ray representation and its systematic study is very informative as it reveals the deep relationship between the representation itself and the group being represented. Before we begin to explore this freedom, however, let us specify the analysis to the case of interest. Let $\{U'_r=\phi(r)U_r\}$ be an admissible set of representatives defined in a suitable neighborhood of $e$. The strong continuity of $U'_r$ and $U_r$ naturally implies that $\phi(r)$ is a continuous complex unimodular function. On the other hand, starting from $\phi(r)$ continuous and $U_r$ strongly continuous, one arrives at $U'_r$ strongly continuous\footnote{Technically additional care should be take in the neighborhood's definition. In fact, being $\mathfrak{N}$ and $\mathfrak{N}'$ the neighborhoods of $e$ where $U'_r$ and $U_r$ are, respectively, defined, then the relation $U'_r=\phi(r)U_r$ takes place in a neighborhood $\mathfrak{K}$ of $e$ such that $\mathfrak{K}\subset (\mathfrak{N}\cap\mathfrak{N}')$. Besides, for admissible representative sets, Eq. (\ref{b1}) is valid for $\phi(r)$ defined in $\mathfrak{K}$ also encompassing products as $rs$.}.  
\medbreak
{\bf Proposition 1:} {\it Let $\omega$ be a local factor defined in a given neighborhood $\mathfrak{N}$ of $e$ and let $\phi(r)$ be a continuous  unimodular complex function such that $\phi(e)=1$ and the function $\omega'$ is given as in (\ref{b1}). Then the function $\omega'$ is also a local factor.} 
\medbreak
{\bf Proof of Proposition 1:} As noted earlier, 1) the strong continuity of admissible representatives implies that $\phi(r)$ is a continuous function. Therefore, $\omega'$ also satisfies (\ref{n8}) for admissible $U'_r$. Moreover, 2) $\omega'(r,e)=\omega'(e,s)=1=\omega'(e,e)$ as can be easily seen. Finally, 3) using (\ref{b1}), it is possible to write
\begin{eqnarray}
\omega'(r,s)\omega'(rs,m)=\omega(r,s)\omega(rs,m)\frac{\phi(r)\phi(s)\phi(m)}{\phi(rsm)} \label{b2}
\end{eqnarray} and using (\ref{n4}) we have 
\begin{eqnarray}
\omega'(r,s)\omega'(rs,m)=\omega(s,m)\omega(r,sm)\frac{\phi(r)\phi(s)\phi(m)}{\phi(rsm)}. \label{b3}
\end{eqnarray} Multiplying the right-hand side of (\ref{b3}) by $1=\phi(sm)/\phi(sm)$ and rearranging the terms, we obtain 
\begin{equation}
\omega'(r,s)\omega'(rs,m)=\omega(s,m)\frac{\phi(s)\phi(m)}{\phi(sm)}\omega(r,sm)\frac{\phi(r)\phi(sm)}{\phi(rsm)},\label{b4}
\end{equation} showing that (\ref{n4}) then holds also for $\omega'$, leading to a local factor. \hspace{.3cm} $\Box$  
\medbreak
We will henceforth study the freedom in selecting admissible representatives on a more comprehensive basis. To do so, it may be a good time to introduce a notation with which the reader is probably more familiar. Let $\delta(r,s)$ be a strongly continuous real function defined for every group element in a suitable chosen neighborhood, satisfying the condition $\delta(e,e)=0$ and 
\begin{equation}
\delta(r,s)+\delta(rs,m)=\delta(s,m)+\delta(r,sm).\label{b5}
\end{equation} Then it is possible to replace the local factors by the so-called local exponents by $\omega(r,s)=e^{i\delta(r,s)}$ (note that (\ref{b5}) is the counterpart of the group associative law to the local exponents).  
\medbreak
{\bf Proposition 2:} {\it For every local exponent, the relations 1) $\delta(r,e)=0=\delta(e,s)$ and 2) $\delta(r,r^{-1})=\delta(r^{-1},r)$ hold.}
\medbreak
{\bf Proof of Proposition 2:} For 1) just take $s=m=e$ ($r=s=e$) in (\ref{b5}) and recall that $\delta(e,e)=0$ to get $\delta(r,e)=0$ ($\delta(e,s)=0$). For 2) take $s=r^{-1}$ in (\ref{b5}) and obtain 
\begin{equation}
\delta(r,r^{-1})+\delta(rr^{-1},m)=\delta(r^{-1},m)+\delta(r,r^{-1}m)\label{b6}
\end{equation} and then take $m=r$ to obtain $\delta(r,r^{-1})=\delta(r^{-1},r)$. \hspace{.3cm} $\Box$

When studying of physical representations, it is relevant to understand when (and how) projective representations may be discarded and one can work directly with genuine representations. This is a subtle aspect, whose answer may go beyond non-trivial points of mathematical theory. We will here appreciate only some of the aspects that are somehow directly related to continuity. In any case, the following definition is crucial for a proper exposition of this topic. 
\medbreak
{\bf Definition 2:} {\it Let $x(r)$ be a continuous real function defined in a neighborhood $\mathfrak{K}$  that includes products of group elements and let $\delta$ and $\delta'$ be local exponents defined in $\mathfrak{N}$ and $\mathfrak{N}'$, respectively. Let it be assumed that $\mathfrak{K}\subset (\mathfrak{N}\cap\mathfrak{N}')$. The local exponents $\delta$ and $\delta'$ will be called equivalent if the relation}
{\it \begin{eqnarray}  
\delta'(r,s)=\delta(r,s)+\Delta_{r,s}[x], \label{b7}
\end{eqnarray} with 
\begin{eqnarray}
\Delta_{r,s}[x]=x(r)+x(s)-x(rs)\label{b8}
\end{eqnarray} holds in $\mathfrak{K}$.}
\medbreak
From Definition 2 above, it is easy to see that $x(e)=0$ insofar as $\delta'(e,e)=\delta(e,e)$. The functional form of $\Delta_{r,s}[x]$ may be, perhaps, better justified by noting from (\ref{b1}) that two equivalent local exponents uniquely define two equivalent local factors with $\phi(r)=e^{ix(r)}$. By slightly changing the order of exposition, it is possible to  enunciate the following proposition.    
 \medbreak
{\bf Proposition 3:} {\it If $\delta$ is a local exponent defined in a given neighborhood, and $x(r)$ a continuous real function such that $x(e)=0$ defined in a suitable neighborhood (see Definition 2), then $\delta'(r,s)$ as defined by (\ref{b7}) and (\ref{b8}) is a local exponent.}    
\medbreak
{\bf Proof of Proposition 3:} First, note that the continuity of $\delta$ and $x$ guarantees continuity for $\delta'$. Moreover, since $x(e)=0$, then $\delta'(e)=0$ directly. Besides 
\begin{eqnarray}
\delta'(r,s)+\delta'(r,sm)=\delta(r,s)+\delta(rs,m)+\Delta_{r,s}[x]+\Delta_{rs,m}[x]\label{b9}
\end{eqnarray} and by means of (\ref{b5}) 
\begin{eqnarray}
\delta'(r,s)+\delta'(r,sm)=\delta(s,m)+\delta(r,sm)+\Delta_{r,s}[x]+\Delta_{rs,m}[x].\label{b10}
\end{eqnarray} Using (\ref{b7}) we now arrive at 
\begin{eqnarray}
\delta'(r,s)+\delta'(r,sm)=\delta'(s,m)+\delta'(r,sm)+\Delta_{r,s}[x]+\Delta_{rs,m}[x]-\Delta_{s,m}[x]-\Delta_{r,sm}[x].
\label{b11}
\end{eqnarray} Finally note that the sum of $\Delta$'s vanishes identically.  \hspace{.3cm} $\Box$
\medbreak
Before proceeding to study the consequences of continuity, we will make some complementary observations. Let us denote the equivalence between two local exponents by $\delta'\sim \delta$. This equivalence relation is 

i) Symmetric: $\delta'\sim \delta$ means $\delta'(r,s)=\delta(r,s)+\Delta_{r,s}[x]$. Therefore $\delta\sim \delta'$ as $x\rightarrow -x$;

ii) Reflexive: obviously $\delta'\sim\delta'$;

iii) Transitive: suppose $\delta_1(r,s)=\delta(r,s)+\Delta_{r,s}[x_1]$ in $\mathfrak{N}_1$ and $\delta_2(r,s)=\delta_1(r,s)+\Delta_{r,s}[x_2]$ in $\mathfrak{N}_2$. Then $\delta_2(r,s)=\delta(r,s)+\Delta_{r,s}[x_1+x_2]$ in some $\mathfrak{N}\subset (\mathfrak{N}_1\cap\mathfrak{N}_2)$, and hence $\delta_2\sim\delta$; Hence (\ref{b7}) and (\ref{b8}) set a formal equivalence class, a truly equivalence class indeed.
  
The observation of the last paragraph can be complemented by the following remark: if $\delta_1$ and $\delta_2$ are local exponents in $\mathfrak{N}_1$ and $\mathfrak{N}_2$ respectively, then every linear combination $\kappa_1\delta_1(r,s)+\kappa_2\delta_2(r,s)$ with $\kappa_i\in\mathbb{R}$ $(i=1,2)$ is also a local exponent in $\mathfrak{N}\subset (\mathfrak{N}_1\cap\mathfrak{N}_2)$.  Summarizing these observations, we  can claim that the equivalence class of a linear combination depends only on the equivalence classes of local exponents entering in the linear combination.     

 This is a good point to appreciate an important theorem due to Weyl \cite{weyl}. 
\medbreak
{\bf Theorem 3:} {\it For a finite dimensional continuous ray representation $\textbf{U}_r$ of a group $H$, every local factor is equivalent to 1.}
\medbreak
{\bf Proof of Theorem 3:} Starting from $U_rU_s=\omega(r,s)U_{rs}=e^{i\delta(r,s)}U_{rs}$, assuming that $n$ is the  dimension of the representation space, and taking the determinant of the above expression, we are left with 
\begin{equation}
\det{U_r}\det{U_s}=e^{in\delta(r,s)}\det{U_{rs}}.\label{b12}
\end{equation} Now,  since $U_r$ is strongly continuous, it follows that $\det{U_r}$ is a continuous function of $r$. Also, $|\det{U_r}|=1=\det{U_e}$.  Thus in a suitable neighborhood, it is possible to write $\det{U_r}\equiv e^{i\Sigma(r)}$ with continuous real functions such that $\Sigma(e)=0$. In this vein, we have from (\ref{b12}), 
\begin{equation}
\delta(r,s)-\frac{\Sigma(r)}{n}-\frac{\Sigma(s)}{n}+\frac{\Sigma(rs)}{n}=0\label{b13}
\end{equation} and since we recognize $x(r)=-\Sigma(r)/n$ we have $\delta(r,s)+\Delta_{r,s}[x=-\Delta/n]=0\sim \delta'(r,s)$, from which $\omega(r,s)\sim 1$ follows. \hspace{.3cm} $\Box$    
\medbreak
This is a  remarkable result,  which makes it clear that continuity of representatives acting in Hilbert spaces is  not only a pleasant and desirable  property but can also constrain important aspects of the representation  that would otherwise  be unspecified.  However, two crucial limitations of the previous result should be noted: first, it is a local achievement, which for this reason is valid in a given neighborhood $\mathfrak{N}$ of $e$. To extend such claim to all group manifold, we should be able to demonstrate it for $\mathfrak{N}=H$. Notably, this is only possible if the zeroth and first homotopy groups of the manifold associated to $H$, $\pi_0(H)$ and $\pi_1(H)$, are both trivial\footnote{It should be mentioned that in order to extrapolate local results for the whole group it is also necessary the group to be compact. This characteristic is, however, fulfilled by every Lie group \cite{haal}.} \cite{nss}. When these requirements are not  satisfied, some additional subtitles may appear,  such as in the case of representations up to a sign for $SO(3)=SU(2)/\mathbb{Z}_2$ rotations,  which are of great impact in spinorial representations. The second point to  emphasize is that the last theorem deals with finite-dimensional representations. We will  continue the analysis  by lifting this  restriction and further investigating continuous representations on general basis.  

As a typical representation, the operators of continuous operator rays form a group under multiplication. This concept  can be systematized by  introducting of the so-called local group $L$,  which entails the freedom in selecting a given operator within a ray and  formalizes, so to speak, the observation with which we started this section. To introduce this group, note that an operator belonging to an admissible set of representatives defined in a suitable neighborhood is given by $e^{i\sigma}U_r$, where the representation continuity  fixes the range of $\sigma$ as the real numbers. Hence 
\begin{eqnarray}
(e^{i\sigma_1}U_r)(e^{i\sigma_2}U_s)=e^{i(\sigma_1+\sigma_2)}U_rU_s=e^{i(\sigma_1+\sigma_2)}\omega(r,s)U_{rs},
\end{eqnarray} or, in terms of local exponents, 
\begin{equation}
(e^{i\sigma_1}U_r)(e^{i\sigma_2}U_s)=e^{i(\sigma_1+\sigma_2+\delta(r,s))}U_{rs}. \label{b14}
\end{equation}  We call $\mathfrak{N}^2$ the neighborhood  comprising the products of any two group elements belonging to $\mathfrak{N}$, and  require that $\mathfrak{N}^2$ is a neighborhood of $e$,  and define $L$ to be the set of elements of the form $[\sigma,r]$ with $\sigma\in\mathbb{R}$ and $r\in\mathfrak{N}$.   
   
 Now define a product $\diamond:L\times L\rightarrow L$ such that
\begin{eqnarray}
[\sigma_1,r]\diamond[\sigma_2,s]=[\sigma_1+\sigma_2+\delta(r,s),rs],\label{n15}
\end{eqnarray} where $\delta(r,s)$ is  a local exponent, in  full agreement with (\ref{b14}) but bypassing any allusion to a given representative (provided it is admissible). 
\medbreak
{\bf Proposition 4:} {\it L is a group under $\diamond$.}
\medbreak
{\bf Proof of Proposition 4:} 1) As it can be readily seen, the unity element is simply given by $[0,e]$; 2)  Note that 
\begin{eqnarray}
[\sigma_1,r_1]\diamond \Big([\sigma_2,r_2]\diamond [\sigma_3,r_3]\Big)=[\theta_1+\theta_2+\theta_3+\delta(r_2,r_3)+\delta(r_1,r_2r_3),r_1r_2r_3],\label{b15}
\end{eqnarray} while 
\begin{eqnarray}
\Big([\sigma_1,r_1]\diamond[\sigma_2,r_2]\Big)\diamond [\sigma_3,r_3]=[\theta_1+\theta_2+\theta_3+\delta(r_1,r_2)+\delta(r_1r_2,r_3),r_1r_2r_3].\label{b16}
\end{eqnarray}  Thus, the associative aspect of the representation inherited by local factors (\ref{b5})  yields an associative product. 3) For every element of $L$, the unity element is reachable by a composition with $[\sigma,r]^{-1}\equiv[-(\sigma+\delta(r,r^{-1})),r^{-1}]$, where $r^{-1}\in\mathfrak{N}$, for
\begin{eqnarray}
 [\sigma,r]\diamond[\sigma,r]^{-1}&=&\left.[\sigma,r]\diamond [-(\sigma+\delta(r,r^{-1})),r^{-1}]=[\sigma-\sigma-\delta(r,r^{-1})+\delta(r,r^{-1}),rr^{-1}]\right.\nonumber\\&=&\left.[0,e]=[\sigma,r]^{-1}\diamond[\sigma,r].\right.\label{b17}
\end{eqnarray} This  concludes the proof. \hspace{.3cm} $\Box$  
\medbreak
Then, the group $L$ has $\mathbb{R}\times\mathfrak{N}^2$ as associated manifold and it is often said that $L$ is the local group constructed for the local exponent $\delta$. Despite the apparent simplicity of the local group, its structure is relevant enough to be analyzed further. Take elements of the form $[\sigma,e]$ of $L$. Of course, these elements form an  one-parameter subgroup, say $C$, of $L$. Let us  present some properties of $C$ that can be easily checked. First, $C$ belongs to the center of $L$. In fact $[\sigma,e]\diamond [\alpha,r]=[\sigma+\alpha+\delta(e,r),er]$, but $\delta(e,r)=0=\delta(r,e)$ since it is a local exponent and of course $er=re=r$. Thus $[\sigma,e]\diamond [\alpha,r]=[\alpha,r]\diamond [\sigma,e]$, $\forall$ $[\alpha,r]$ $\in L$. Now  it is not hard to prove that $[\sigma,r]=[\sigma,e]\diamond [0,r]$ and then every element of $L$  can be written in terms of an element belonging to $C$ (something relevant in what follows) together with an element of $H$.  

The central group investigation is important to understand a relevant aspect between the center of algebras and the local existence of projective representations. In fact, by inspecting elements of $C$ one sees that it comprises all the information about the freedom in a ray selection. To see that the inspection of $L$ (and therefore $C$) is indeed informative about representations of $H$, let us show some relevant isomorphisms. 
\medbreak
{\bf Lemma 2:} {\it The quotient group $L/C$ is locally isomorphic to H.}
\medbreak
{\bf Proof of Lemma 2:} As known, the group $L/C$ has elements belonging to the set $\{\kappa C| \kappa\in L\}$, while $H$ comprises elements $r_1,r_2,\ldots$. Let $\varphi$ be an application from $L/C\rightarrow H$. Before  setting $\varphi$ completely, we note that $\varphi(\kappa C)$ typically has $\varphi([\sigma_1,r]\diamond [\sigma_2,e])$ as arguments. It turns out that  
 
\begin{eqnarray}
[\sigma_1,r]\diamond [\sigma_2,e]=[0,r]\diamond [\sigma_1,e]\diamond [\sigma_2,e]=[0,r]\diamond [\sigma_1+\sigma_2,e].\label{b18} 
\end{eqnarray} Then calling $\sigma_1+\sigma_2\equiv\sigma$ we have $[\sigma_1,r]\diamond [\sigma_2,e]=[0,r]\diamond [\sigma,e]$. The definition of $\varphi$ is thus completed by selecting the pure $H$ element of its domain, that is $\varphi([0,r]\diamond [\sigma,e])\doteq r$. In view of this we see that 1) $\varphi(e)=\varphi([0,e]\diamond[0,e])=\varphi([0,e])$ and hence $\varphi([0,e])\doteq e=\varphi(e)$. Now note that   
\begin{equation}\label{b19}
\varphi\Big([0,r]^{-1}\diamond [\sigma,e]^{-1}\Big)=\varphi\Big([-\delta(r,r^{-1}),r^{-1}]\diamond [-(\sigma+\delta(e,e)),e]\Big),
\end{equation} where we have used $e^{-1}=e$. Since $\delta(e,r)=0$, $\forall r$ locally, we have 
\begin{equation}\label{b20}
\varphi\Big([0,r]^{-1}\diamond [\sigma,e]^{-1}\Big)=\varphi\Big([-\delta(r,r^{-1}), r^{-1}]\diamond [-\sigma,e]\Big)=\varphi\Big([-\sigma-\delta(r,r^{-1}),r^{-1}]\Big)\doteq r^{-1}.
\end{equation}  Finally, since $\varphi([0,r]\diamond [\sigma,e])=r$, we arrive at $\varphi(g^{-1})=\varphi^{-1}(g)$ $\forall g\in L/C$. \hspace{.3cm} $\Box$ 
 \medbreak
Incidentally,  we might note that $\varphi([\sigma_1,e]\diamond[0,r_1]\diamond[\sigma_2,e]\diamond[0,r_2])=\varphi([\sigma_1+\sigma_2+\delta(r_1,r_2),r_1r_2])\doteq r_1r_2$. However, obviously, $\varphi([\sigma_i,e]\diamond [0,r_i])\doteq r_i$ ($i=1,2$) and then $\varphi(g_1g_2)=\varphi(g_1)\varphi(g_2)$, $\forall g_1,g_2, g_1g_2 \in L/C$.  

Besides the important result just described in Lemma 2, it is also possible to locally connect local groups  with equivalent exponents. This is the content of the next  lemma. 
\medbreak
{\bf Lemma 3:} {\it Let $\delta$ and $\tilde{\delta}$ be two equivalent local exponents in a given neighborhood, that is $\tilde{\delta}(r,s)=\delta(r,s)+\Delta_{rs}[x]$. Then the corresponding local groups $L$ and $\tilde{L}$ are locally isomorphic.}
\medbreak
{\bf Proof of Lemma 3:} Consider the mapping $\varphi:L\rightarrow \tilde{L}$ such that $\varphi([\sigma,r])=[\sigma-x(r),\tilde{r}=r]$. It is clear that 
\begin{eqnarray}
\varphi([\sigma_1,r_1])\diamond\varphi([\sigma_2,r_2])=[\sigma_1-x(r_1),r_1]\diamond [\sigma_2-x(r_2),r_2]=[\sigma_1+\sigma_2-x(r_1)-x(r_2)+\tilde{\delta}(r_1,r_2), r_1r_2].\label{b21}
\end{eqnarray} On the other hand, it can be readily verified that 
\begin{equation}
\varphi([\sigma_1,r_1\diamond [\sigma_2,r_2])=\varphi([\sigma_1+\sigma_2+\delta(r_1,r_2),r_1r_2])=[\sigma_1+\sigma_2+\delta(r_1,r_2)-x(r_1r_2),r_1r_2].\label{b22} 
\end{equation}  Thus, the equality of (\ref{b21}) and (\ref{b22}) follows directly from the equivalence of the local exponents. \hspace{.3cm} $\Box$   
\medbreak
For completeness, we mention another general result concerning local isomorphisms. If $z$ is a real  nonzero constant and the local groups $L$ and $L'$ are constructed for $\delta$ and $\delta'=z\delta$, respectively, then the mapping $f:L\rightarrow L'$ such that $f([\sigma,r])=[\sigma'=z\sigma,r'=r]$ defines an isomorphism between $L$ and $L'$. In fact, $f([\sigma_1,r_1])\diamond f([\sigma_2,r_2])=[z\sigma_1,r_1]\diamond[z\sigma_2,r_2]=[z\sigma_1+z\sigma_2+\delta'(r_1,r_2),r_1r_2]$. By its turn $f([\sigma_1,r_1]\diamond[\sigma_2,r_2])=f([\sigma_1+\sigma_2+\delta(r_1,r_2),r_1r_2])=[z(\sigma_1+\sigma_2+\delta(r_1,r_2)),r_1r_2]$ and, since $\delta'=z\delta$, $L\simeq L'$ locally.  

Now we can  resume the discussion around Eq. (\ref{b17}). As Lemma 3 asserts,  studying isomorphism between local groups is a  way to study local exponents. Moreover, Lemma 2 ensures that locally $L/C\simeq H$, that is $L$ is the extension of $H$ by $C$.  Taken together, these two results point to the center of $L$, the $C$ group, as the  really relevant group to  consider for the study of local exponents per se. The Lie algebra of the group $L$ is at the heart of the question of whether a given representation is projective or genuine \cite{gilm}. With effect, local exponents are (locally) equivalent to zero provided  that a rearrangement of algebra generators removes  their central counterpart. The  study of continuity allows  us to consider this result here from the group perspective: it is clear from the formulation that removing local exponents is possible if the group $C$ whose elements are given by $[\sigma,e]$ can be ruled out from the analysis. This observation  ultimately leads to the famous Bargmann's theorem,  which states that central charges of semi-simple Lie algebras  can always be excluded by redefining the algebra generators \cite{bar}.      

\section{Final Remarks}

We have discussed some aspects of Wigner's approach to dealing with (irreducible) representations of the Poincar\'e group. All the perspectives presented here focused on the rigor that Wigner devoted to the concept of representation's continuity itself. Indeed, this formal aspect has of course profound implications for the foundations of representation theory and, as we tried to make clear in the manuscript, its consequences are also relevant to physics. 

In revisiting the underlying concepts for constructing the representation, we made a special effort to evoke all the relevant steps along with the proofs, and to provide a path through the concepts whose fulcrum rested on the matter of continuity. In this way we were able to connect in advance (algebraic and topological) aspects of representation theory in physics, as well as some discussion about the projective or genuine representations issue, to the primitive idea of selecting admissible representatives.

At the risk of exaggerating somewhat, we would say that the formal consideration of some aspects of continuous representation in Hilbert spaces is perhaps one of the main ingredients of Wigner's success in finding a formal characterization of particles, along with other equally relevant aspects in the formulation, such as the exploration of induced representations \cite{mac,mactwo} and finding of new representations. 

The rigorous effort of Wigner's approach to formalize aspects previously bypassed by others is better explained by a well-known comment due to Wigner himself \cite{comm}:  

\begin{quote} 
The mathematical formulation of the physicist’s often crude experience leads in an uncanny number of cases to an amazingly accurate description of a large class of phenomena. This shows that the mathematical language has more to commend it than being the only language which we can speak; it shows that it is, in a very real sense, the correct language.
\end{quote}

Just over 80 years have passed since the work celebrated here and, as befits of a well-made work that survives the scrutiny of time, its fundamental concepts and approaches are still precise, elegant, and revealing insightful perspectives.  

\begin{appendices}

\section{Appendix: Symmetry Representations}

For completeness, we present here a step-by-step proof of the famous Wigner theorem of 1931 on symmetry operations in Hilbert space \cite{wig1}. More often than never, the proof of this theorem is revisited in several contexts and with different levels of sophistication (see \cite{two} for a complete list of references and two quite interesting different proofs). Bargmann presented an elegant version of the proof \cite{bw}, while a deep understanding of the Wigner viewpoints, together with some discussion about other proofs, is given in Ref. \cite{che}. The presentation given here follows the steps performed by Weinberg \cite{wei} due to its completeness. After that, we also prove that for an identity component subgroup, every action is unitary. Let us start by contextualizing this rather important result. 

We say that two descriptions of a given quantum mechanical system are isomorphic if there is a one-to-one correspondence, $\bm{\Psi}\leftrightarrow\bm{\Psi'}$, between the rays describing the physical system preserving probabilities, i.e., $\bm{\Phi}\cdot\bm{\Psi}=\bm{\Phi'}\cdot\bm{\Psi'}$. This is the desired situation for the description of the quantum system in two inertial reference frames connected by Lorentz transformations, for instance. Moreover, transformations that preserve the ray internal product (and hence the probabilities) are called symmetry transformations. As is well known, the Wigner theorem to be invoked here shows that every isomorphic ray correspondence engenders a (also one-to-one) vector correspondence in Hilbert space as $\psi'=U\psi$ and the general properties of $U$ are listed in the theorem statement.

{\bf Theorem 4:} {\it Every symmetry transformation in the Hilbert space of physical states can be represented by an operator that is either linear and unitary or anti-linear and anti-unitary.}

{\bf Proof of Theorem 4:} Let $\mathfrak{R}_1\supset \psi_1$ and $\mathfrak{R}_2 \supset \psi_2$ be two rays in the Hilbert space $\mathcal{H}$ and, analogously, let $\mathfrak{R}'_1\supset \psi'_1$ and $\mathfrak{R}'_2\supset \psi'_2$ for transformed rays (and corresponding elements). Let $\varphi \in End(\mathcal{H})$ transforming $\mathfrak{R}$ into $\mathfrak{R}'$ be a symmetry transformation such that $|(\psi_1,\psi_2)|^2=|(\psi'_1,\psi'_2)|^2$. In addition, we will also require the existence of $\varphi^{-1}$ as a symmetry transformation. Denoting by $\{\psi_k\}\in \mathfrak{R}$ a complete orthonormal set of states, it is easy to see that orthonormality is inherited for $\{\psi'_k\}\in \mathfrak{R}'$, obtained by means of $\varphi$. Indeed, since $|(\psi_m,\psi_n)|^2=\delta_{mn}$, we have $|(\psi'_m,\psi'_n)|^2=\delta_{mn}$, from which  
\begin{eqnarray}
(\psi'_m,\psi'_n)(\psi'_m,\psi'_n)^*=\delta_{mn}.\label{ap1}
\end{eqnarray} In this context, we note that although a notation tailored to the results for finite dimensional spaces has been used, the adaptation to general cases is straightforward. In accordance with more precise aspects (as completeness), the validation steps will be performed with no concern to a particularization for finite dimensions.    

Let us consider Eq. (\ref{ap1}) for the $m=n$ case. Let $k$ be such that $(\psi'_m,\psi'_m)=k$. As it is well known (a quantum mechanical postulate) $0\leq (\psi'_m,\psi'_m) \in \mathbb{R}$ (the equality holding for the null state). Therefore, Eq. (\ref{ap1}) implies $k^2=1$, from which  $(\psi'_m,\psi'_m)=1$ necessarily. Consider now the case $m\neq n$ and take $(\psi'_m,\psi'_n)=a+ib$, where $a,b \in \mathbb{R}$. In this case Eq. (\ref{ap1}) translates to $a^2+b^2=0$, leading to $a=0=b$ and hence $(\psi'_m,\psi'_n)=0$ for $n\neq m$. Therefore, $(\psi'_m,\psi'_n)=\delta_{mn}$ and the transformation $\varphi$ does preserve orthonormality. 

In accounting for completeness, suppose $\{\psi'_k\}$ is not a complete set. Then, there must exist a given state, say $\tilde{\psi}'$ which has not projection in the set (or a subset of) $\{\psi'_k\}$, that is $(\tilde{\psi}',\psi'_k)=0$, $\forall k$. This fact would imply $|(\tilde{\psi}',\psi'_k)|^2=0$ and, via $\varphi^{-1}$, this would lead to $|(\tilde{\psi}',\psi'_k)|^2=|(\tilde{\psi},\psi_k)|=0$, a clear contradiction since $\{\psi_k\}$ is complete. Therefore the set $\{\psi'_k\}$ is also complete (and orthonormal) and a given basis of the original set is transformed into a basis in the arriving set. 

Take now an element of the set $\{\psi_k\}$, say $\psi_{\bar{k}}$, and define $\phi_k=\frac{1}{\sqrt{2}}(\psi_{\bar{k}}+\psi_k)$ for $k\neq\bar{k}$. The element $\phi_k$ belongs to $\mathfrak{R}$. Notice, from the previous definition, that 
\begin{eqnarray} 
|(\psi_{\bar{k}}, \phi_k)|^2=\frac{1}{2}\big|(\psi_{\bar{k}},\psi_{\bar{k}})+(\psi_{\bar{k}},\psi_k)\big|^2=\frac{1}{2}.\label{apn1}
\end{eqnarray} Besides, as we showed the transformation to preserve completeness, any vector belonging to $\mathfrak{R}'$ may be written as 
\begin{equation}
\phi'_k=\sum_n \alpha_{kn}\psi'_n,\label{ap2}
\end{equation} from which we have 
\begin{eqnarray}
|(\psi'_{\bar{k}},\phi'_k)|^2=\Big|\sum_n \alpha_{kn}(\psi'_{\bar{k}},\psi'_n)\Big|^2=|\alpha_{k\bar{k}}|^2.\label{ap3}
\end{eqnarray} For a symmetry transformation, Eqs. (\ref{apn1}) and (\ref{ap3}) should be equal, leading to $|\alpha_{k\bar{k}}|=\frac{1}{\sqrt{2}}$. This reasoning may be repeated for $|(\psi_k,\phi_k)|^2$ and $|(\psi'_k,\phi'_k)|^2$ resulting in $|\alpha_{kk}|=\frac{1}{\sqrt{2}}$. Also, again repeating the same procedure, a bit of simple algebra shows that the coefficients $\alpha_{km}$ all vanish for $k\neq \bar{k}$ and $k\neq m$ (with $m\neq \bar{k}$). The first part of the demonstration is completed by choosing\footnote{This choice, as it should be clear in the course of the proof, does not change the conclusions.} $\alpha_{k\bar{k}}=1/\sqrt{2}=\alpha_{kk}$ and calling the transformation engendering it by $U$, in such a way that 
\begin{equation}
\phi'_{k}=U\phi_k=\frac{1}{\sqrt{2}}(U\psi_{\bar{k}}+U\psi_k). \label{ap4}
\end{equation} The task now is to extend the founded symmetry transformation to all Hilbert space. 

Let us now take $\psi=\sum_m \beta_m\psi_m$ from which we have 
\begin{equation}
|(\psi_k,\psi)|^2=\Big|\sum_m \beta_m(\psi_k,\psi_m)\Big|^2=|\beta_k|^2.\label{ap5}
\end{equation} For a $U$ transformed vector $\psi'=\sum_m \beta'_m U\psi_m$ it can be readily verified that $|(\psi'_k,\psi')|^2=|\beta'_k|^2$ and, therefore, $|\beta_k|^2=|\beta'_k|^2$, $\forall$ $k$. In particular, $|\beta_{\bar{k}}|^2=|\beta'_{\bar{k}}|^2$ and hence $|\beta_k|^2/|\beta_{\bar{k}}|^2=|\beta'_k|^2/|\beta'_{\bar{k}}|^2$ or simply\footnote{In the case that $\beta_1$ is null, one should particularize another $k$ and perform the same analysis. Since the $\psi$ state is assumed to exists, eventually a given $k$ would lead to a non vanishing coefficient.}
\begin{equation}
\Bigg|\frac{\beta_k}{\beta_{\bar{k}}}\Bigg|^2=\Bigg|\frac{\beta'_k}{\beta'_{\bar{k}}}\Bigg|^2.\label{ap6}
\end{equation} Taking into account $\phi_k$ and $\psi$ we have 
\begin{equation}
|(\phi_k,\psi)|^2=\Bigg|\Bigg(\frac{1}{\sqrt{2}}[\psi_{\bar{k}}+\psi_k],\sum_m \beta_m\psi_m\Bigg)\Bigg|^2=\frac{1}{2}|\beta_{\bar{k}}+\beta_k|^2.\label{ap7}
\end{equation} Analogously
\begin{equation}
|(\phi'_k,\psi')|^2=\Bigg|\Bigg(\frac{1}{\sqrt{2}}[U\psi_{\bar{k}}+U\psi_k],\sum_m \beta'_m U\psi_m\Bigg)\Bigg|^2=\frac{1}{2}|\beta'_{\bar{k}}+\beta'_k|^2,\label{ap8}
\end{equation} allowing one to write $|\beta_{\bar{k}}+\beta_k|^2=|\beta'_{\bar{k}}+\beta'_k|^2$. This last relation implies $|\beta'_{\bar{k}}+\beta'_k|^2/|\beta'_{\bar{k}}|^2=|\beta_{\bar{k}}+\beta_k|^2/|\beta'_{\bar{k}}|^2$ and as $|\beta_1|^2=|\beta'_1|^2$ we have $|\beta'_{\bar{k}}+\beta'_k|^2/|\beta'_{\bar{k}}|^2=|\beta_{\bar{k}}+\beta_k|^2/|\beta_{\bar{k}}|^2$, or equivalently 
\begin{equation}
\Bigg|1+\frac{\beta'_k}{\beta'_{\bar{k}}}\Bigg|^2=\Bigg|1+\frac{\beta_k}{\beta_{\bar{k}}}\Bigg|^2.\label{ap9}
\end{equation} 

Calling for a moment $\beta_k/\beta_{\bar{k}}=x+iy$ (and similarly $\beta'_k/\beta'_{\bar{k}}=x'+iy'$), Eqs. (\ref{ap6}) and (\ref{ap9}) provide, respectively, 
\begin{equation}
x^2+y^2=x'^2+y'^2\label{ap10}
\end{equation} and
\begin{equation}
(1+x)^2+y^2=(1+x')^2+y'^2\label{ap11}
\end{equation} whose combination demands $x=x'$ and $y=\pm y'$, that is 
\begin{eqnarray}  
Re\Bigg(\frac{\beta'_k}{\beta'_{\bar{k}}}\Bigg)=Re\Bigg(\frac{\beta_k}{\beta_{\bar{k}}}\Bigg),\nonumber\\
Im\Bigg(\frac{\beta'_k}{\beta'_{\bar{k}}}\Bigg)=\pm Im\Bigg(\frac{\beta_k}{\beta_{\bar{k}}}\Bigg).\label{ap12}
\end{eqnarray} These are the constraints to be taken in extending the $U$ operation to all Hilbert space. By choosing the upper sign in Eq. (\ref{ap12}) we have $\beta'_k/\beta'_{\bar{k}}=\beta_k/\beta_{\bar{k}}$, while the down sign implies $\beta'_k/\beta'_{\bar{k}}=\beta^*_k/\beta^*_{\bar{k}}$. 

An important complement, due to Weinberg, to the original proof is the demonstration that the choice  $\beta'_k/\beta'_{\bar{k}}=\beta_k/\beta_{\bar{k}}$ or $\beta'_k/\beta'_{\bar{k}}=\beta^*_k/\beta^*_{\bar{k}}$ must be made for all cases. To see this is indeed the case, suppose that for some $k$ we have $\beta'_k/\beta'_{\bar{k}}=\beta_k/\beta_{\bar{k}}$, while for $m\neq k$ we have instead $\beta'^*_m/\beta'^*_{\bar{k}}=\beta_m/\beta_{\bar{k}}$. Define now a normalized $\chi$ such that 
\begin{equation}
\chi=\frac{1}{\sqrt{3}}(\psi_{\bar{k}}+\psi_k+\psi_m),\label{ap13}
\end{equation} for which, of course, $\chi'=U\chi$. Repeating again the previous procedure being used so far, it is not difficult to see that 
\begin{eqnarray}
|(\chi,\psi)|^2=\Bigg|\Bigg(\frac{1}{\sqrt{3}}(\psi_{\bar{k}}+\psi_k+\psi_m),\sum_n \beta_n\psi_n\Bigg)\Bigg|^2=\frac{1}{3}|\beta_{\bar{k}}+\beta_k+\beta_m|^2,\label{ap14}
\end{eqnarray} while $|(\chi',\psi')|^2=|\beta'_{\bar{k}}+\beta'_k+\beta'_m|^2/3$, leading to $|\beta_{\bar{k}}+\beta_k+\beta_m|^2=|\beta'_{\bar{k}}+\beta'_k+\beta'_m|^2$. Recalling that $|\beta_k|^2=|\beta'_k|^2$, $\forall$ $k$, we have 
\begin{eqnarray}
\frac{|\beta_{\bar{k}}+\beta_k+\beta_m|^2}{|\beta_{\bar{k}}|^2}=\frac{|\beta'_{\bar{k}}+\beta'_k+\beta'_m|^2}{|\beta'_{\bar{k}}|^2}.\label{ap15}
\end{eqnarray} Taking into account the supposition assumed within this paragraph, Eq. (\ref{ap15}) may be recast in the form 
\begin{equation}
\Bigg|1+\frac{\beta_k}{\beta_{\bar{k}}}+\frac{\beta^*_m}{\beta^*_{\bar{k}}}\Bigg|^2=\Bigg|1+\frac{\beta_k}{\beta_{\bar{k}}}+\frac{\beta_m}{\beta_{\bar{k}}}\Bigg|^2.\label{ap16}
\end{equation} A bit of algebra shows that the constraint presented in Eq. (\ref{ap16}) translates to 
\begin{equation}
\frac{\beta_m\beta_k}{\beta_{\bar{k}}\beta_{\bar{k}}}+\frac{\beta^*_m\beta^*_k}{\beta^*_{\bar{k}}\beta^*_{\bar{k}}}=\frac{\beta^*_m\beta_k}{\beta^*_{\bar{k}}\beta_{\bar{k}}}+\frac{\beta_m\beta^*_k}{\beta_{\bar{k}}\beta^*_{\bar{k}}},\label{ap17}
\end{equation} or simply 
\begin{equation}
Re\Bigg(\frac{\beta_k\beta^*_m}{\beta_{\bar{k}}\beta^*_{\bar{k}}}\Bigg)=Re\Bigg(\frac{\beta_k\beta_m}{\beta_{\bar{k}}\beta_{\bar{k}}}\Bigg).\label{ap18}
\end{equation} Finally, calling $\beta_k/\beta_{\bar{k}}\equiv u=u_1+iu_2$ and $\beta_m/\beta_{\bar{k}}\equiv v=v_1+iv_2$, the above equation ($Re(uv^*)=Re(uv)$) can only be satisfied if $u_2v_2=0$, that is 
\begin{equation}
Im\Bigg(\frac{\beta_k}{\beta_{\bar{k}}}\Bigg)Im\Bigg(\frac{\beta_m}{\beta_{\bar{k}}}\Bigg)=0,\label{ap19}
\end{equation} a clear contradiction regarding a complex space of states. Therefore, one is left with either $\beta'_k/\beta'_{\bar{k}}=\beta_k/\beta_{\bar{k}}$ or $\beta'_k/\beta'_{\bar{k}}=\beta^*_k/\beta^*_{\bar{k}}$. In the case $\beta'_k=\beta_k$, $\forall$ $k$, then
\begin{equation}
U\Big(\sum_k \beta_k\psi_k\Big)=\sum_k \beta_k U\psi_k,\label{ap20}
\end{equation} or else ($\beta'_k=\beta^*_k$, $\forall$ $k$)
\begin{equation}
U\Big(\sum_k \beta_k\psi_k\Big)=\sum_k \beta^*_k U\psi_k.\label{ap21}
\end{equation}

It is important to have a clear-cut of what was demonstrated. Equations (\ref{ap20}) and (\ref{ap21}) show that a same choice must necessarily be made for element inside a given vector state. The theorem demonstration is finalized by showing that it is impossible for a given same transformation that a given vector state transform like (\ref{ap20}) and others as (\ref{ap21}) dictates. Taking advantage of the finite dimension spaces notation, let $\mathbb{V}\subset \mathcal{H}$ be a Hilbert subspace spanned by $\{\psi_k\}$ and $k=1,\ldots,l,\ldots,\dim{\mathbb{V}}$. Suppose the existence of $\varphi$ for which $\varphi'=\sum_k a_k U\psi_k$ and $\eta$ such that $\eta'=\sum_k b^*_k U\psi_k$ for the same operator $U$. Within this context, a vector $\rho=\sum_{k=1}^{l-1}a_k\psi_k+\sum_{k=l}^{\dim{\mathbb{V}}}b_k\psi_k$ would transform under $U$ so as to violate the result encoded in (\ref{ap20}) and (\ref{ap21}). The finalization of the proof now follows straightforwardly. Let $\psi_1=\sum_k \alpha_k\psi_k$ and $\psi_2=\sum_k \beta_k\psi_k$. Assuming $U$ acting according to (\ref{ap20}), i.e. linearly, we have
\begin{eqnarray} 
(U\psi_1,U\psi_2)=\sum_{m,n}\alpha^*_m\beta_n(U\psi_m,U\psi_n)=\sum_{m,n}\alpha^*_m\beta_n\delta_{mn}=(\psi_1,\psi_2),\label{ap22}
\end{eqnarray} asserting $U$ as unitary. By instead demanding (\ref{ap21}), its anti-linearity imply  
\begin{eqnarray} 
(U\psi_1,U\psi_2)=\sum_{m,n}\alpha_m\beta^*_n(U\psi_m,U\psi_n)=\sum_{m,n}\alpha_m\beta^*_n\delta_{mn}=(\psi_2,\psi_1)=(\psi_1,\psi_2)^*,\label{ap23}
\end{eqnarray} from which an anti-unitary action may be read. \hspace{.3cm} $\Box$ 
\medbreak
We finalize this appendix by recalling that for the largest connected subgroup (the identity subgroup), the orthochronous proper Lorentz subgroup, for example, every operator ray is unitary.  
\medbreak
{\bf Proposition 5:} {\it Let $H$ be a connected group (or equivalently an identity\footnote{A technicality should be mentioned here: in general, an identity component refers to the largest connected set in the manifold associated to the group containing the identity element $e$ of the group.} (sub)group). Then $\textbf{U}_r$ is unitary for all $r\in H$.} 
\medbreak
{\bf Proof of Proposition 5:} In a suitable neighborhood $\mathfrak{N}\in H$ containing $e$, every group element $r$ can be written as $r=s^2$ ($s$ being, of course, another group element). By the theorem just exposed, operators acting upon the Hilbert space $\mathcal{H}$ are either linear and unitary or anti-linear and anti-unitary, and hence an isomorphic ray correspondence defines ray operators endowed with the same properties. It turns out, however, that the square of a unitary or anti-unitary operator is unitary. To see that, recall that if $\textbf{U}_s$ is anti-linear, then $\textbf{U}_r=\textbf{U}^2_s$ and for any $\phi, \psi \in \mathcal{H}$ it follows
\begin{eqnarray}
(\textbf{U}_r\phi,\textbf{U}_r\psi)=(\textbf{U}_s[\textbf{U}_s\phi],\textbf{U}_s[\textbf{U}_s\psi])=(\textbf{U}_s\psi,\textbf{U}_s\phi)=(\phi,\psi).\label{ul}
\end{eqnarray} In the case of $\textbf{U}_s$ the proof follows straightforwardly\footnote{Note, in addition, that from the isometry condition $\textbf{U}_r^\dagger\textbf{U}_r=1$ (and the coisometry condition $\textbf{U}_r\textbf{U}^\dagger_r=1$ as well) this property is readily obtained.}. 

As $H$ is connected, every element $r\in H$ equals a given finite product, say $r=\prod_{i=1}^n r_i$, of elements in $\mathfrak{N}$. Hence $\textbf{U}_r=\prod_{i=1}^n \textbf{U}_{r_{i}}$. Now, for every $\textbf{U}_{r_{i}}$ take a partition, as before, in such a way that
\begin{equation}
\textbf{U}_{r}=\prod_{i=1}^n\textbf{U}^2_{s_{i}}.
\end{equation} Then, the ray operator $\textbf{U}_r$ is given by the a finite product of unitary (ray) operators, from which it is seen that $\textbf{U}_r$ is itself unitary.  \hspace{.3cm} $\Box$ 
\medbreak
\end{appendices}
 
\section*{Acknowledgments}

JMHS thanks to CNPq (grant No. 303561/2018-1) for financial support. GMCR thanks to CAPES for financial support.

\end{document}